# Designing the Metaverse: A Scoping Review to Map Current Research Effort on Ethical Implications

**Matteo Zallio[1], Takumi Ohashi[2], and P. John Clarkson[1]**

[1] University of Cambridge, Department of Engineering, Engineering Design Centre Cambridge, CB2 1PZ, United Kingdom

[2] Tokyo Institute of Technology, 2-12-1 Ookayama, Meguro-ku, Tokyo 152-8550, Japan

**ABSTRACT**

The metaverse and digital, virtual environments have been part of recent history as places in which people can socialize, work and spend time playing games. However, the infancy of the development of these digital, virtual environments brings some challenges that are still not fully depicted. With this article, we seek to identify and map the currently available knowledge and scientific effort to discover what principles, guidelines, laws, policies, and practices are currently in place to allow for the design of digital, virtual environments, and the metaverse. Through a scoping review, we aimed to systematically survey the existing literature and discern gaps in knowledge within the domain of metaverse research from sociological, anthropological, cultural, and experiential perspectives. The objective of this review was twofold: (1) to examine the focus of the literature studying the metaverse from various angles and (2) to formulate a research agenda for the design and development of ethical digital, virtual environments. With this paper, we identified several works and articles detailing experiments and research on the design of digital, virtual environments and metaverses. We found an increased number of publications in the year 2022. This finding, together with the fact that only a few articles were focused on the domain of ethics, culture and society shows that there is still a vast amount of work to be done to create awareness, principles and policies that could help to design safe, secure and inclusive digital, virtual environments and metaverses.

# INTRODUCTION

The metaverse is not a recent concept. In 1992 Neal Stephenson wrote a science fiction novel, Snow Crash, in which the term Metaverse appeared for the first time (Stephenson, 1992).

Later on in 2003, Linden Labs created Second Life, a virtual environment in which people could create avatars and immerse themselves in a digital life with other individuals and recreate with a different vision what could be a copy of the physical world (Linden Labs, 2003; Malaby, 2009).

Following these early attempts to build digital immersive environments, investments from the gaming industry, including Roblox, Active Worlds, Epic Games and many other businesses, boosted several opportunities for the metaverse and its correlated applications to be deployed at a wider scale (Businesswire, 2023).

Around the end of the second decade of the 2000, other tech businesses approached the metaverse. In 2021 the social media company Facebook pivoted its brand to Meta Platforms, as a way to "bring together Facebook apps and technologies under one new company brand and focus on bringing the Metaverse to life by helping people connect, find communities and grow businesses" (Facebook, 2021).

NVIDIA with the Omniverse started to promote an extensible open platform built for virtual collaboration and real-time accurate simulation for creators to connect design tools, assets, and projects to collaborate and iterate in a shared virtual space (Shapiro, 2021).

With all of these and many more advances, the first few years of the third decade of the 2000s can be seen as the immersive technology renaissance, where the metaverse can be broadly defined as a set of multiple digital, virtual environments developed and owned by multiple companies (Zallio & Clarkson, 2022; Evans, 2018).

Businesses, experts, developers, designers, researchers are in the process of unfolding several layers of complexity around the study and the creation of different digital worlds which will have impactful implications on the behavioral, sociological and psychological aspects of human beings (Zallio & Clarkson, 2022).

With the advent of these technologies and its implications the need to understand what research has been done so far to envision, develop and deploy principles, guidelines, laws, policies, and practices to allow for the design of digital, virtual environments and metaverses becomes increasingly important (Zallio & Clarkson, 2022).

With this article, we aim to identify and map the currently available knowledge and scientific effort done so far in this field. Through a scoping review, we systematically surveyed the existing literature and discerned gaps in knowledge from sociological, anthropological, cultural, and experiential perspectives. The objective of this review was twofold: (1) to examine the focus of the literature studying the metaverse from various angles and (2) to formulate a research agenda for the design and development of ethical digital, virtual environments.

**RESEARCH METHODOLOGY**

To examine the literature studying the metaverse from various angles and to formulate a research agenda for the design and development of a better metaverse we run a scoping review according to five phases.

We first identified relevant studies that focused on metaverse research from different angles, with a focus on sociological, anthropological, cultural, and experiential perspectives. We then developed different inclusion and exclusion criteria and we included peer-reviewed studies with a focus on the metaverse, and digital, virtual environments, published in English between 2015 and 2021. In the third step we extracted data such as author, year of publication, abstract and main findings from the studies that met the inclusion criteria. The fourth step focused on performing and in-depth review of the studies from a qualitative perspective as described in Table 1. The fifth step encompassed the data synthesis phase done by categorizing the studies based on their focus and by examining the main findings from each study. The results of the data synthesis were used to formulate a discussion around what potential research agenda for the design and development of a better metaverse could be created.

The scoping review was executed in accordance with the PRISMA-ScR method (Tricco, et al., 2018) and conducted on November 4th, 2022, utilizing the Web of Science All Database, which encompasses multiple databases, such as the Web of Science Core Collection, Current Contents Connect, Derwent Innovations Index, KCI-Korean Journal Database, MEDLINE, Data Citation Index, BIOSIS Citation Index, and SciELO Citation Index. The search was performed using the keyword "metaverse" and was restricted to open-access, peer-reviewed articles written in English.

The review excluded public reports, Articles that do not contain the word "metaverse" in the title or abstract, Articles considering the metaverse as one of the applications. (e.g., R&D of a specific technology that can contribute to the metaverse), Articles with only abstract or conference proceedings. The screening criteria, as described in Table 1, were established to ensure that only articles relevant to the research questions were analyzed.

**Table 1.** An overview of the screening criteria.

| No. | Inclusion criteria | Exclusion criteria |
|---|---|---|
| 1 | Full articles. | Public reports. |
| 2. | English written articles. | Articles that do not contain the word "metaverse" in the title or abstract. |
| 3. | Open-access articles. | Articles considering the metaverse as one of the applications. (e.g., R&D of a specific technology that can contribute to the metaverse). |
| 4. | Articles dealing with sociological, anthropological, cultural, and experiential aspects of the metaverse. | Articles with only abstract or conference proceedings. |

Initially, original papers were extracted from the collection of articles obtained through a search query for "metaverse" in the topic field using the Web of Science search engine. Subsequently, only open-access articles written in English were extracted.

The screening process was conducted by excluding articles that did not feature the word "metaverse" in the title and abstract. The first and second authors conducted the initial screening independently based on the criteria outlined in Table 1. In cases where there was disagreement between the authors, a consensus was reached through a further qualitative and comprehensive analysis in real time. The second author then thoroughly examined the full text to eliminate any remaining articles that failed to meet the screening criteria specified in Table 1.

This study adopted an interpretive worldview and utilized a generative coding approach (Eakin & Gladstone, 2020) to map previous research on the metaverse.

The generative coding was executed by the second author, who read each article's title, abstract, and full text to identify the central concepts addressed. The generated codes were then reviewed, synthesized, and merged into second-order codes. Finally, these second-order codes were grouped into categories that represent the prevailing themes addressed in prior metaverse research.

## FINDINGS

Figure 1 shows the PRISMA-ScR flowchart of the scoping review process. The Web of Science All Databases yielded a total of 1365 articles, which were then reduced to 181 by selecting original, English-language, open-access journal articles.

Of these 181 papers, those that did not mention the word "metaverse" in the title or abstract were excluded, resulting in a pool of 155 articles. The first and second authors independently reviewed the titles and abstracts of these papers to identify those that focused on the metaverse. In cases where the authors disagreed on the inclusion or exclusion of an article, consensus was reached through a further qualitative and comprehensive analysis in real time, resulting in a total of 75 articles.

An additional 6 articles were excluded based on full-text screening, including 1 article due to lack of full access, 3 articles that were considered commentary, 1 article as a protocol paper, and 1 conference article. The final pool of 69 research articles were subjected to a comprehensive literature review in this study.

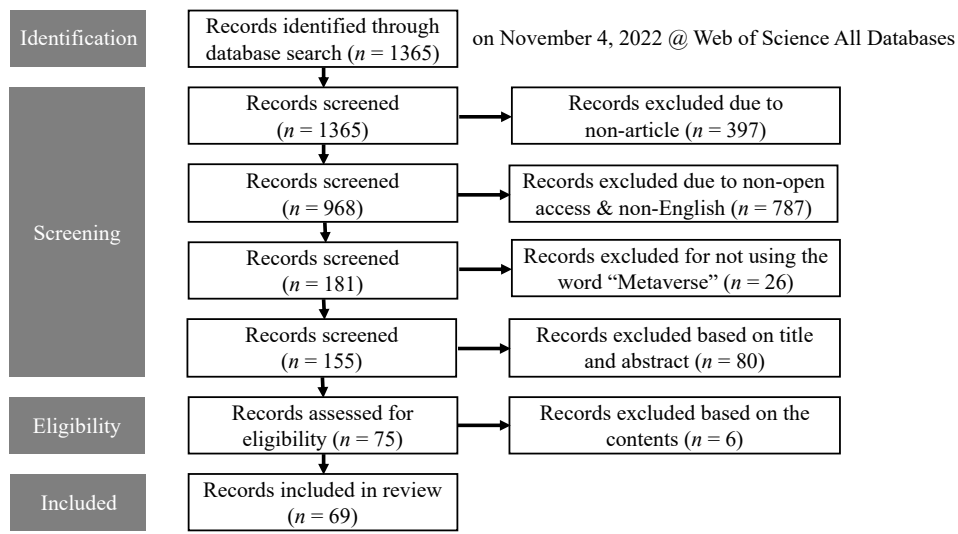

**Figure. 1:** A schematic of the PRISMA-ScR flowchart.

The 69 selected research articles were analyzed by the second author using a generative coding method. The analysis process resulted in the identification of 12 second-order codes that encapsulated the central themes addressed in the metaverse literature.

These codes included the following themes: theoretical/conceptual frameworks, user acceptance, healthcare, education, technology, specific applications, smart city, mental/psychological impact, tourism, sociological/cultural discussion, ethics, and user experience.

The results of the generative coding process were tabulated and synthesized in Table 2 which also grouped the themes into five categories.

**Table 2.** Categories identified during the scoping review.

| Category | Second-order code | Description | # of article |
|---|---|---|---|
| **Technology applications** | Healthcare, education, specific applications, smart city, tourism | Applied research using metaverse-related technologies, especially in the areas of healthcare, education, smart cities, and tourism. | 28 |
| **Usability, user acceptance and user experience** | User acceptance, cognitive/psychological impacts, user experience | Research on the experience of users in the metaverse, including the perception and acceptance of the metaverse and its physical/cognitive effects. | 14 |
| **Technology development** | Technology | Research on technologies for developing the metaverse, including digital twin, including digital technologies such as blockchain, VR/AR, and technologies related to security. | 10 |
| **Ethics, culture and society** | Sociological/cultural discussion, ethics | Research discussing the ethical, cultural, and social implications and prospects of the emerging metaverse. | 9 |
| **Theoretical and conceptual framework** | Theoretical/conceptual frameworks | Research providing theoretical/conceptual frameworks and or analysis of the current state and prospects in relation to metaverse research using existing theoretical frameworks. | 8 |

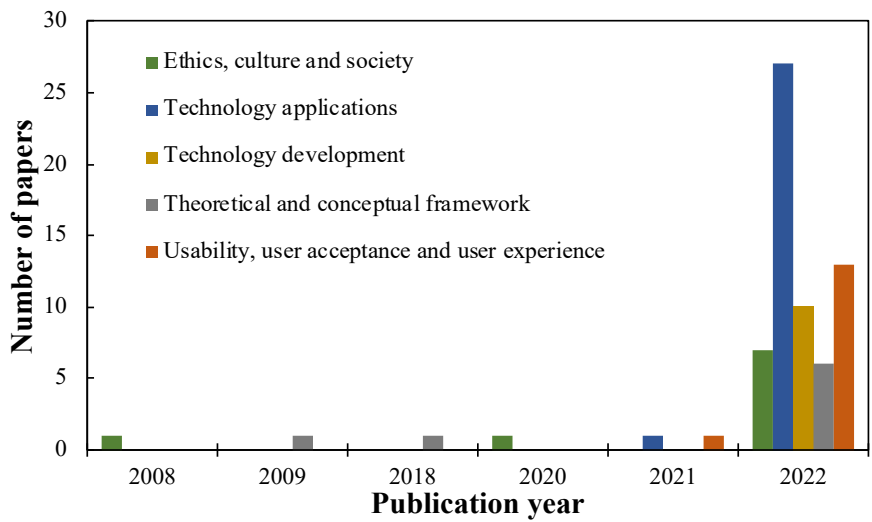

**Figure 2:** Publication trends identified through the scoping review.

Additionally, the analysis of the distribution of the number of publications in each category based on the year of publication is showed in Figure 2. It is evident that prior to 2021, the number of metaverse publications was quite limited, with only 6 publications. However, in 2022 there was a significant increase in the number of metaverse-related publications.

## DISCUSSION

The research findings show that there has been a growing body of literature focusing on the metaverse and its various aspects in recent years. The flowchart developed after the analysis of data from the scoping review illustrates the systematic approach used to identify and select relevant research articles for this study. The causes that generated these results can be various: the hyped use of the word metaverse, the focus of this work and the databases used to perform the search, the impact that certain technologies are starting to have on their users, among many other causes.

The results of the selection process showed that the pool of selected articles was reduced from 1365 to 69 through a series of filters, including language, access, and relevance criteria in order to have a better and more focused view of the aspects related to the five main categories identified.

The analysis of the selected 69 research articles showed that there are several key categories that are addressed in the metaverse literature, including theoretical and conceptual frameworks, user acceptance, healthcare, education, technology, specific applications, smart city, mental and psychological impact, tourism, sociological and cultural discussion, ethics, and user experience. These categories were aimed to provide a deeper understanding of the state of knowledge in the field of metaverse research.

One of the interesting observations from the study is the significant increase in the number of metaverse-related publications in 2022, compared to the limited number of publications prior to 2021. The reasons behind this surge might be several, and some hypotheses are the increased interest in the topic due to a hyped use of the term metaverse, the recognition from the scientific community that the field is still quite unexplored, or the need from organizations to learn more about the challenges and opportunities of using such new technologies. These are some of the many hypotheses to further investigate with more research on different areas spanning ethics, technology, behavioural science and other fields.

An interesting aspect that emerges from this research is the low number of studies classified under the category ethics, culture and society and theoretical and conceptual framework such as the one developed by Dahan et al. (Dahan et al., 2022).

This early finding might suggest an interpretation which leads to identify a lack of research and a consequent opportunity to develop more research effort and studies around topics of ethics, integrity, security, inclusion, accessibility related to digital, virtual environments.

Furthermore this work highlights that there is limited literature on both ethics, culture, and society and theoretical and conceptual frameworks. Given this scarcity, by looking at some previous works (Zallio & Clarkson, 2022 a, b), which presented early results depicted through the lens of an ethical viewpoint on the metaverse, they offer ideas to develop novel conceptual frameworks for future assessments of the metaverse. Notwithstanding this early attempts, we still have several opportunities to grow these fields in which the study, analysis, assessment of digital, virtual environments and metaverses from an ethical, integrity, security, accessibility and inclusion point of view can lead to develop better, more comprehensive frameworks, principles and guidelines for the future.

Although some early attempts, such as the field of Metavethics (Zallio & Clarkson, 2023), have been made to establish new areas of study that could influence the design and development of digital and virtual environments and metaverses, we are still at a very early age with little effort developed to create disciplines to study topics regarding inclusion, safety, security and access of digital, virtual environments and the metaverse.

## CONCLUSION AND FUTURE WORK

With this article, we aimed to explore with an initial scoping review the scenario of research work done in relation to the metaverse and digital, virtual environments focusing on different aspects of the technology applications, the usability and user acceptance, the technology development, the ethics, culture and society and the theoretical and conceptual frameworks.

This scoping review highlighted that there has been a recent increase in the number of metaverse-related publications, with a majority of the research focused on technology and specific applications. However, there is still a significant gap in knowledge and research developed in relation to ethics, culture, and society. This is a crucial aspect to ensure the design of safe, secure, and inclusive digital, virtual environments, and metaverses.

It is important to note that the limitations of this study include the fact that it was based on a limited number of articles that were available in open-access

English-language journals and the results may not necessarily reflect the entire body of research on the metaverse and digital, virtual environments.

Furthermore, the results are based on the analysis performed through lenses that might be leading to focus on certain aspects of the research, rather than the totality. Therefore, future studies should aim to expand the scope of the literature search to include other databases and languages and to involve multiple researchers in the analysis process.

In terms of future work, further research is needed to fill in the gaps in knowledge identified in this scoping review. This could include more in-depth studies of the ethical, cultural, and societal implications of the metaverse, as well as the development of new disciplines and fields of research such as the Metavethics, the ethics applied to the metaverse and digital, virtual environments, but also develop principles, policies, and practices to guide the design and development of digital, virtual environments that provide value to their users.

In conclusion, the research findings provide initial, valuable insights into the state of knowledge in the field of metaverse research and highlight the key themes and areas of focus in this field for the future.

## ACKNOWLEDGMENT

Dr. Matteo Zallio and Prof. Takumi Ohashi equally contributed to the development and writing of this article. Prof. P John Clarkson contributed to writing the conclusions and future work section. This study was supported by the Sustainable Ethics for Inclusive Digital Environments (SEIDE) initiative at the Digital Education Futures Initiative at Hughes Hall, University of Cambridge, the Metavethics Institute, and the Engineering Design Centre, Inclusive Design Group from the University of Cambridge.

## REFERENCES

Businesswire, (2023). Global Metaverse Market Report 2023: Size, Trends and Forecasts 2022-2027 with Profiles of Top Players - Epic Games, Meta Platforms, Microsoft, Inworld AI, NetEase, Nvidia, and Roblox : https://www.businesswire.com/news/home/20230124005875/en/Global-Metaverse-Market-Report-2023-Size-Trends-and-Forecasts-2022-2027-with-Profiles-of-Top-Players---Epic-Games-Meta-Platforms-Microsoft-Inworld-AI-NetEase-Nvidia-and-Roblox---ResearchAndMarkets.com, (Accessed 16 February 2023).

Dahan, N.A.; Al-Razgan, M.; Al-Laith, A.; Alsoufi, M.A.; Al-Asaly, M.S.; Alfakih, T. (2022). Metaverse Framework: A Case Study on E-Learning Environment (ELEM). *Electronics* 2022, *11*, 1616. https://doi.org/10.3390/electronics11101616

Eakin, J. M., & Gladstone, B. (2020). "Value-adding" Analysis: Doing More With Qualitative Data. International Journal of Qualitative Methods, 19. https://doi.org/10.1177/1609406920949333

Evans, L. (2018). The Re-Emergence of Virtual Reality (1st ed.). Routledge. https://doi.org/10.4324/9781351009324

Facebook, 2021. Introducing Meta: A Social Technology Company. https://about.fb.com/news/2021/10/facebook-company-is-now-meta/. (Accessed 16 February 2023).

Linden Labs. (2003).  Retrieved February 09 from https://www.lindenlab.com/about

T. Malaby, 2009. *Making virtual worlds: Linden Lab and Second Life*. Ithaca, N.Y.: Cornell University Press.

Shapiro, E., 2021. The Metaverse Is Coming. Nvidia CEO Jensen Huang on the Fusion of Virtual and Physical Worlds. Time: https://time.com/5955412/artificial- intelligence-nvidia-jensen-huang/. (Accessed 5 April 2022).

Stephenson, N. (1992). Snow crash. Publisher Bantam Books.

Tricco, A. C, Lillie, E., Zarin W., et al. (2018). PRISMA Extension for Scoping Reviews (PRISMA-ScR): Checklist and Explanation. Ann Intern Med.2018;169:467-473. doi:10.7326/M18-0850

Zallio, M., & Clarkson, P. J. (2022 a). Designing the Metaverse: A study on Inclusion, Diversity, Equity, Accessibility and Safety for digital immersive environments. Telematics and Informatics, 101909. https://doi.org/https://doi.org/10.1016/j.tele.2022.101909

Zallio, M., Clarkson, P. J. (2022 b). Inclusive Metaverse. How businesses can maximize opportunities to deliver an accessible, inclusive, safe Metaverse that guarantees equity and diversity. Technical report ENG-TR.013. University of Cambridge. ISSN: 2633-6839. https://doi.org/10.17863/CAM.82281

Zallio, M., Clarkson, P. (2023). Metavethics: Ethical, integrity and social implications of the metaverse. In: Tareq Ahram, Waldemar Karwowski, Pepetto Di Bucchianico, Redha Taiar, Luca Casarotto and Pietro Costa (eds) Intelligent Human Systems Integration (IHSI 2023): Integrating People and Intelligent Systems. AHFE (2023) International Conference. AHFE Open Access, vol 69. AHFE International, USA. http://doi.org/http://doi.org/10.54941/ahfe1002891